\documentclass[12pt]{article}
\usepackage{a4}
\usepackage{psfig}
\usepackage{epsfig}
\usepackage{epsf}
\usepackage{rotating}
\usepackage{citesort}       
\usepackage{amssymb}
\usepackage{graphicx}

\newcommand{\lsim}{\mbox{$\leq$}}

\newcommand{\anueR}{\mbox{$\overline{\nu}_{eR}$}}

\newcommand{\text}[1]{\mbox{$\rm #1 $}}

\def\plb#1#2#3{    { Phys. Lett. }{\bf B #1} (19#2) #3}

\begin{document}
\begin{flushright}
{\small
FTUAM-03-314\\
}
\end{flushright}

\vspace{0.2cm}
\begin{center}
{ \Large \bf  KamLAND Bounds on Solar Antineutrinos and 
neutrino transition magnetic moments}\\[0.2cm]
{\large  E.~Torrente-Lujan$^{a\star}$\\[2mm]
$^a$ {\small\sl Dept. Fisica Teorica C-XI,\\ 
Univ. Autonoma de Madrid, 
28049 Madrid, Spain,}\\
}


\end{center}

\abstract{
We investigate  the possibility of detecting 
solar electron antineutrinos with the KamLAND experiment. 
These electron antineutrinos  are 
predicted by spin-flavor oscillations at a significant rate 
even if this mechanism is not the leading solution to the 
SNP. 

KamLAND is sensitive to antineutrinos originated  
from solar ${}^8$B neutrinos.
From KamLAND  negative results after 145 days of data taking,
 we obtain model independent limits on 
 the total flux of solar electron antineutrinos
$\Phi( {}^8 B)< 1.1-3.5\times 10^4\ cm^{-2}\ s^{-1}$,
 more than one order of magnitude smaller than existing
 limits,
and on their appearance probability 
$P<0.15\%$ (95\% CL).

Assuming a concrete model for 
antineutrino production by spin-flavor precession,
 this upper bound   implies an 
upper limit on the product of the intrinsic neutrino magnetic 
moment and the value of the solar magnetic
  field $\mu B< 2.3\times 10^{-21}$ MeV 
95\% CL (for LMA $(\Delta m^2, \tan^2\theta)$ values).

Limits on neutrino transition moments are also obtained. 
For realistic values of other astrophysical solar
 parameters these upper 
limits would imply that the neutrino magnetic 
moment is constrained to be, in the most conservative case, 
$\mu\lsim 3.9\times 10^{-12}\ \mu_B$ (95\% CL) for a relatively small
field $B= 50$ kG.  
For  higher values of the magnetic field we obtain:
 $\mu\lsim 9.0\times 10^{-13}\ \mu_B$  for 
field $B= 200$ kG and 
 $\mu\lsim 2.0\times 10^{-13}\ \mu_B$  for 
field $B= 1000$ kG at the same statistical significance.

}

\vskip .5truecm

{PACS: }

\vfill
{\small {}$^\star$ e-mail:  emilio.torrente-lujan@cern.ch.

Present address: Dept. Fisica, GFT. Universidad de Murcia. 
Murcia. Spain.
}


\newpage

\section{Introduction}
Evidence of eelctron antineutrino disappearance in a beam of antineutrinos 
in the KamLAND experiment has been recently 
presented \cite{kloctober}.
The analysis of these results \cite{kloctober,klothers} 
in terms of neutrino oscillations
have   largely 
improved our knowledge of neutrino mixing in the LMA region.
The results appear to confirm in a independent way that the observed 
deficit of solar neutrinos is indeed due to neutrino oscillations. 
The ability to measure the LMA solution, the one
preferred by the solar neutrino data at present,
 ``in the lab'' puts KamLAND in a pioneering situation: after these 
results there should remain little doubt of the physical reality of 
neutrino mass and oscillations.
Once neutrino mass is observed,
neutrino magnetic moments are an inevitable consequence in the 
Standard Model and beyond. Magnetic moment interactions 
arise in any renormalizable gauge theory only as finite 
radiative corrections: the diagrams which contribute to the
neutrino mass will inevitably generate a magnetic moment
once the external photon line is added.

The KamLAND experiment is the 
successor of previous reactor experiments (
CHOOZ~\cite{CHOOZ}, PaloVerde~\cite{PaloVerde}) 
at a much larger scale in 
terms of baseline distance and total incident flux.  
This experiment relies upon a 1 kton liquid scintillator
 detector   located at the old, enlarged,  Kamiokande site.
 It searches for the oscillation of antineutrinos 
emitted by several nuclear power plants in Japan. 
The nearby 16  (of a total of 51) nuclear power stations deliver 
a $\overline{\nu}_e$ flux of $1.3\times 10^6 cm^{-2}s^{-1}$
for neutrino energies $E_\nu>1.8$ MeV at the detector position. 
About $85\%$ of this flux comes from  reactors forming a 
well defined baseline of 139-344 km. Thus, the flight range 
is limited  in spite of using  several reactors, because of this 
fact the sensitivity of KamLAND   increases by nearly two 
orders of magnitude compared to previous reactor experiments.

Beyond reactor neutrino measurements, the  secondary 
physics program 
of KamLAND includes diverse objectives as the measurement of geoneutrino flux 
emitted by the radioactivity  of the earth's crust and mantle, the detection 
of antineutrino bursts from galactic supernova and, after extensive  
improvement of the detection sensitivity, the detection of low energy 
${}^7 Be$ neutrinos using neutrino-electron elastic scattering.

Moreover, the KamLAND experiment is  capable of 
detecting  potential electron antineutrinos produced on fly 
from solar ${}^8$B neutrinos \cite{alianiantinu}. 
These  antineutrinos are 
predicted by spin-flavor oscillations at a significant rate if the neutrino is a Majorana particle and 
if its magnetic moment is high enough
\cite{generalrandom,specificrandom}.  
In Ref.\cite{alianiantinu} as been remarked 
 that the flux of reactor antineutrinos at the Kamiokande site 
is comparable, and in fact smaller, to the flux of 
${}^8$B neutrinos emitted by the sun
,$\Phi( {}^8 B)\simeq 5.6\times 10^6 cm^{-2}s^{-1}$ 
\cite{kloctober,bpb2001,Ahmad:2002jz}.
Their energy spectrum  is important at energies 
$2-4$ MeV while solar neutrino spectrum peaks at around 
$9-10$ MeV. As the inverse beta decay reaction cross section
increases as the square of the energy, we would 
expect nearly 10 times 
more solar electron antineutrino events even if the 
 initial fluxes were equal in magnitude.

The publication of the  SNO 
results~\cite{Ahmad:2002ka,Ahmad:2002jz}
has already 
made an important breakthrough towards the solution of the long standing
 solar neutrino 
\cite{Aliani:2002ma,Strumia:2002rv,Aliani:2002er,Aliani:2001zi}
problem (SNP) possible.
These results provide the strongest evidence so 
far (at least until KamLAND 
improves its statistics) for flavor oscillation 
in the neutral lepton sector. 

The existing  bounds on solar electron antineutrinos are 
 strict.
 The present upper limit on the absolute flux of solar antineutrinos
originated from ${}^8 B$ neutrinos is 
\cite{alianiantinu,antibounds,PDG2002}
$\Phi_{\overline{\nu}}({}^8 B)< 1.8\times 10^5\ cm^{-2}\ s^{-1}$
which is equivalent to an averaged conversion probability bound
of $P<3.5\%$ (SSM-BP98 model). There are also bounds on their differential 
energy spectrum \cite{antibounds}: the conversion probability is smaller 
than $8\%$ for all $E_{e,vis}>6.5$ MeV going down the $5\%$ level above 
$E_{e,vis}\simeq 10$ MeV.

The main aim of this work is to study the implications of the 
recent KamLAND results on the determination of the 
solar electron antineutrino appearance probability, independently from 
concrete models on antineutrino production. 
The structure of this work is the following.
In section 2
 we discuss the main features of KamLAND experiment 
that are relevant for our analysis:
The
salient aspects of the procedure we are adopting and 
the  results of our analysis are presented and discussed in
 sections 3. 
In Section 4 we apply the results we obtained in a particular 
model for the solar magnetic field, we obtain bounds on the 
values of the intrinsic neutrino transition magnetic moments.
Finally, 
in section 5 we draw our 
conclusions and discuss possible future scenarios.

\section{The computation of the expected signals}
\label{klsignal}

\subsection{The KamLAND signal}
\label{Kamland}

Electron antineutrinos from any source, nuclear reactors or
solar origin,  with energies 
above 1.8 MeV are measured in KamLAND by detecting the inverse 
$\beta$-decay reaction $\overline{\nu}_e+p\to n+e^+$. The time 
coincidence, the space correlation and the energy balance  
between the positron signal and the 2.2 MeV $\gamma$-ray
 produced by the capture of a already-thermalized  neutron on a
 free proton make it possible to identify this reaction 
unambiguously, even in the presence of a rather large background. 

The two principal ingredients in the calculation of the 
expected 
signal in KamLAND are  the corresponding   flux and
the electron antineutrino cross section on protons.
The average number of positrons $N_i$ originated from the 
solar source which 
are detected per visible energy bin $\Delta E_i$ is given by the convolution 
of different quantities: 
\begin{eqnarray} \hspace{-0.3cm}
N_i&=& Q_0 \int_{\Delta E_i}dE_e \int_0^\infty dE_e^r \epsilon(E_e)R(E_e,E_e^r)
\int_{E_e^r}^\infty dE_{\overline{\nu}} \overline{p}(E_{\overline{\nu}})
\Phi (E_{\overline{\nu}})
 \sigma (E_{\overline{\nu}},E_e^r)
\label{e3466} 
\end{eqnarray}
where $Q_0$ is a normalization constant  accounting for the 
fiducial volume and live time of the experiment,
$\overline{p}$.
Expressions for the 
electron antineutrino capture cross section  
$\sigma (E_{\overline{\nu}},E_e^r)$ 
are taken from 
the literature \cite{vogel,kltorrente}. The  matrix 
element  for this cross section can 
be written in terms of the neutron half-life, 
 we have used the latest published 
value $t_{1/2}=613.9\pm 0.55$ \cite{PDG2002}.
The functions $\epsilon(E_e)$ and $R(E_e,E_e^r)$ are
 the detection efficiency and  the energy 
resolution function.
We use in our analysis  the following
expression for the energy resolution in the prompt 
positron detection
$\sigma(E_e)=0.0062+ 0.065\surd E_e$ .
This expression is obtained from the raw calibration 
data presented in Ref.\cite{klstony}.
Note that we prefer to use this  expression instead of the 
much less accurate one given in 
Ref.\cite{kloctober}.
Moreover, we assume a  408 ton fiducial 
mass and  the detection efficiency is 
taken  independent of the energy \cite{kloctober},
 $\epsilon=80\%$.
In order to obtain concrete limits, a model should  be taken which 
predict $\overline{p}$ and its dependence with the energy.
For our purpose it will suffice  to suppose  $\overline{p}$ a 
constant. This is justified at least in two cases: a) 
if the energy range $\Delta E$ over which we perfom the integration 
is small enough 
so the variation of the probability is not very large, or b) if 
we reinterpret ${\overline p}$ as an energy-averaged probability,
 note that, in a general case, this is always true because the 
un-avoidable convolution with a  finite energy resolution. 
(see Expression 10 in Ref.\cite{specificrandom}):
\begin{eqnarray}
{\overline p}_{\Delta E}&=&  
\int_{\Delta E} dE\ \sigma(E) \Phi(E) P_{\overline \nu}(E)/
{\int_{\Delta E} dE\ \sigma(E) \Phi(E)}.
\label{avnu}
\end{eqnarray}
Let us finally note that independently of the reasons above, 
upper 
limits to be obtained on continuation are still 
valid even if the antineutrino probabilities are significantly 
different from constant: if we take 
${\overline p}=\max_{\Delta E} P_{\overline \nu}(E)$ 
the  expected antineutrino signal computed 
with ${\overline p}$ will 
be always larger than the signal obtained inserting the full 
probability.

\begin{figure}
\centering
\psfig{file=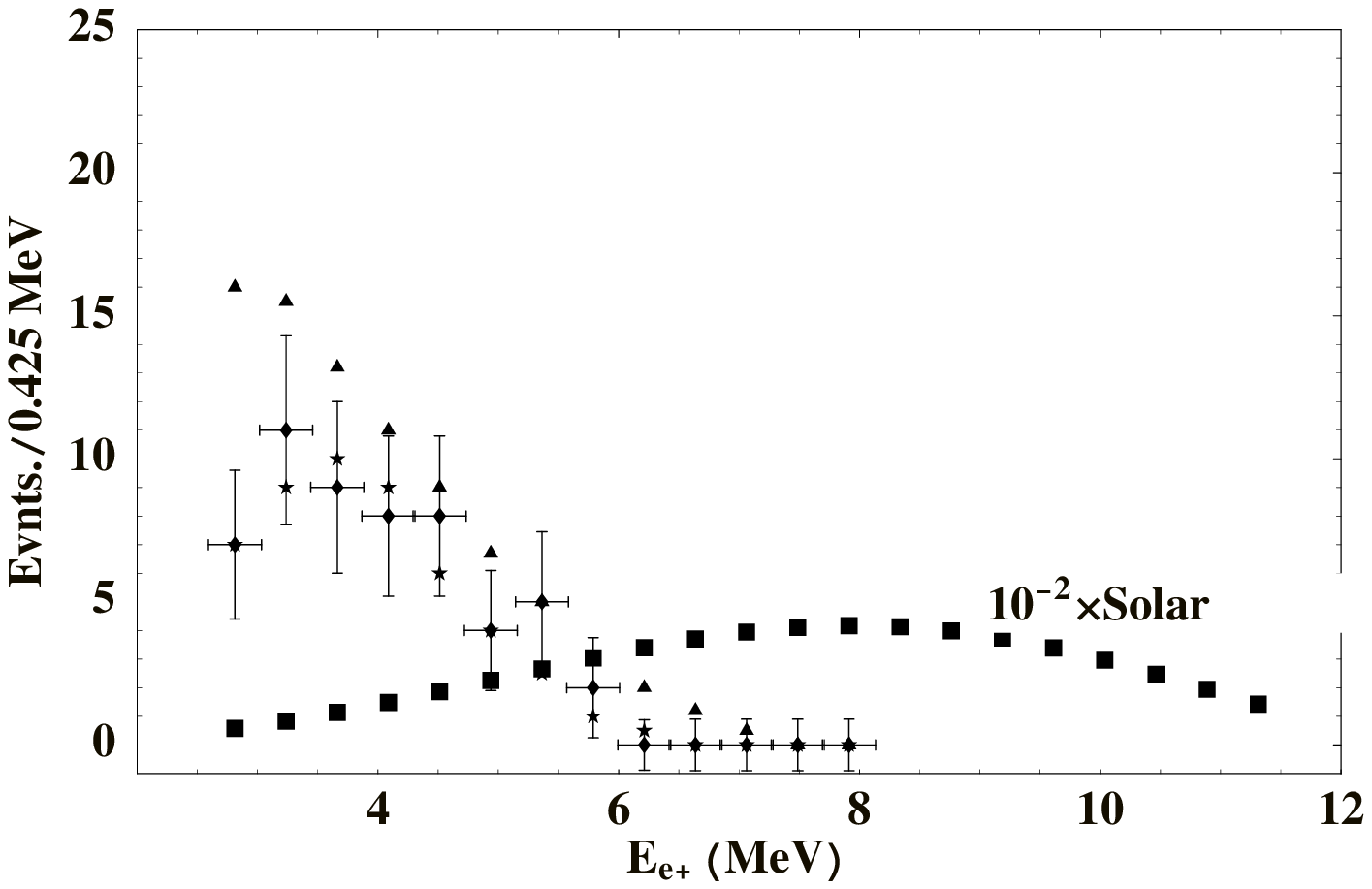,width=11cm} 
\label{f1}
\caption{
The KamLAND  positron spectra from reactor antineutrinos 
(from Fig.5 in Ref.\protect\cite{kloctober}):
 measured (145.5 days), MC expectations in absence of oscillations
and best fit including neutrino oscillations 
($\Delta m^2=6.9\times 10^{-5}$ eV$^2$, $\sin^2\theta=1$,
respectively points with error-bars, triangles and stars).
The ``solar'' positron spectrum  (black solid squares) 
obtained assuming the 
shape of the ${}^8 B$ 
neutrino flux and a total normalization
$10^{-2}\times \Phi({}^8 B)$ (that is, an overall 
$\nu_e-\overline{\nu}_e$ 
conversion probability $\overline{P}\sim 1\%$).}
\end{figure}

The results of our simulation are summarized
 in Fig.\ref{f1} where we show  
the ``solar'' positron spectrum  obtained assuming the 
shape of the ${}^8 B$ 
neutrino flux and a total normalization
$10^{-2}\times \Phi({}^8 B)$ which means an overall 
$\nu_e-\overline{\nu}_e$ 
conversion probability $\overline{P}\sim 1\%$.

In addition we have computed the expected signal coming from 
antineutrino reactors. 
A number of short baseline experiments 
(See~Ref.\cite{Murayama:2000iq,vogel2} and references therein) 
have previously measured the energy spectrum of reactors at distances 
where oscillatory effects have been shown to be inexistent. 
They have shown that the theoretical neutrino flux predictions 
are reliable within 2\% \cite{piepke}.
The effective flux of antineutrinos released by the nuclear
 plants is a rather well  understood function of 
the thermal power of the reactor and
 the amount of thermal power emitted during the 
fission of a given nucleus, which gives the total amount, and 
the  isotopic composition of the reactor fuel which gives the 
spectral shape.  
Detailed tables for these
 magnitudes can be found in 
Ref.~\cite{Murayama:2000iq,vogel2mq}.
For a given isotope the energy  spectrum can be parametrized 
by an exponential expression \cite{vogel}
where the coefficients depend 
on the nature of the fissionable isotope 
(see Ref.\cite{Murayama:2000iq} for explicit values).
Along the year, between periods of refueling, the total 
effective flux changes 
with time as the fuel is expended and the isotope 
relative composition varies.
We take the average of the relative fission yields over the live 
time as given by the experiment: 
${}^{235} U=57\% $,
${}^{238} U=7.8\% $,
${}^{239} Pu=30\% $,
${}^{241} Pu=5.7\% $.
In order to obtain the expected number of events at KamLAND, 
we sum the expectations for all the relevant reactor sources weighting 
each source by its power and distance to the detector
(table II in Ref.~\cite{Murayama:2000iq}), 
 assuming the same 
spectrum originated from each reactor. 
We sum over the nearby  power reactors, we neglect 
farther Japanese and Korean reactors and even farther 
rest-of-the-world reactors which  give only a minor additional
contribution.

\section{Analysis and Results}

We will obtain upper bounds on the solar electron antineutrino 
appearance probability analyzing the observed KamLAND rates
in three different ways. In the first one, we will make a 
standard $\chi^2$ analysis of the observed and expected solar 
signals in the 13 prompt positron energy bins considered by 
KamLAND \cite{kloctober}. 
In the second and third cases we will 
apply Gaussian and poissonian probabilistic 
considerations to the 
global rate seen by the experiment and to the individual 
event content in the
highest energy bins ($E_e> 6 $ MeV) where KamLAND 
observes zero events.
This null signal 
 makes particularly simple the extraction of statistical 
conclusions in this case.

\subsection{ Analysis of the KamLAND Energy Spectrum }

Here we fully 
use the  binned KamLAND signal (see  Fig. 5 in  Ref.\cite{kloctober}) 
for  estimating
 the parameters of solar electron antineutrino production 
from the method of  maximum-likelihood.
We minimize the quantity
\begin{eqnarray}
\chi^2 &=& \chi^2_{i=1,9}+\chi^2_{i=10,13}
\label{chi2}
\end{eqnarray}
where the first term correspond to the contribution of the
first nine bins where the signal is large enough and  
 the use of the
 Gaussian approximation is justified. 
The second term correspond
to the latest bins where the observed and expected signals
are very small and poissonian statistics is needed. 
The explicit expressions are:
\begin{eqnarray}
\chi^2_{i=1,9}&=& \sum_{i=1,9} 
\frac{(S_i^{exp}-S_i^{teo})^2}{\sigma^2}   \\
\chi^2_{i=10,13} &=&2\sum_{i=10,13} 
 S_i^{teo}-S_i^{exp}+ S_i^{exp}\log \frac{S_i^{exp}}{S_i^{teo}}.
\end{eqnarray}
The quantities $S_i$ are the observed bin contents from KamLAND.
The theoretical signals are in principle a function 
of three different parameters: the solar electron antineutrino 
appearance probability $\overline{p}$  and
 the neutrino oscillation parameters $(\Delta m^2,\theta)$.
Both contributions, the contribution from solar antineutrinos 
and that one from solar reactors, can be treated as different
summands:
\begin{eqnarray}
S_i(\overline{p},
\Delta m^2,\theta)&=&
S_i^{solar}(\overline{p})+
S_i^{reactor}(\Delta m^2,\theta).
\end{eqnarray}
According to our model, 
the solar antineutrino 
appearance probability  $\overline{p}$ is  
taken as a constant and we can finally write:
\begin{eqnarray}
S_i(\overline{p},
\Delta m^2,\theta)&=&
\overline{p}\times S_i^{0}+
S_i^{reactor}(\Delta m^2,\theta).
\end{eqnarray}
In this work
we will take for the minimization 
values of the oscillation parameters those obtained 
when ignoring any solar antineutrinos
(LMA solution 
$\Delta m^2=6.9\times 10^{-5}$ eV$^2$, $\sin^2\theta=1$
 from Ref.\cite{kloctober}) and 
we will perform a one-parameter minimization with respect
$\overline{p}$. 
This approximation is well justified because the solar 
antineutrino probability is clearly very small, 
We avoid in this way  the simultaneous 
 minimization 
 with respect to the three parameters 
($\overline{p},  \Delta m^2,\theta)$.

We perform a 
 minimization of the one dimensional  function
 $\chi^2(\overline{p})$. 
to test a particular oscillation hypothesis against the 
parameters of the best fit 
and obtain the allowed interval  in 
$\overline{p}$
parameter space 
taking into account the asymptotic properties of the 
likelihood function, i.e. $\log {\cal L}-\log {\cal L}_{min}$
behaves asymptotically as a $\chi^2$ with one degree of 
freedom.
In our case, the  minimization can be performed analytically because 
of the simple, lineal, dependence. 
A given point in the confidence interval
 is allowed if 
 the globally subtracted quantity 
fulfills the condition 
 $\Delta\chi^2=
\chi^2 (\overline{p})-\chi_{\rm min}^2<\chi^2_n(CL)$.
Where $\chi^2_{n=1}(90\%,95\%,...)=2.70,3.84,..$ are the 
quantiles for one  degree of freedom.

Restricting to physical values 
of $\overline{p}$, the minimum 
of the $\chi^2$ function is obtained for 
$\overline{p}=0$. The corresponding confidence intervals are
$\overline{p}<4.5 \% $ (90\% CL) and 
$\overline{p}<7.0\% $ (95\% CL). 
We have explicitly checked, varying the concrete place
where the division 
between ``Gaussian'' and ``poissonian'' bins is 
established in Expression \ref{chi2},
that the values of these upper limits are 
largely insensitive to details of our analysis.
In particular, similar upper limits are obtained
in the extreme cases: if  Gaussian or poissonian 
statistics is employed for all 13 bins.
These upper limits are considerably weaker than those 
obtained in the next section. One possible  reason for that is that 
they are  obtained applying asymptotic general arguments to the 
$\chi^2$ distribution, stronger, or more precise  
limits could be obtained if a  
Monte Carlo simulation of the distribution of the 
finite sample $\chi^2$ distribution is performed 
(where the boundary condition $\overline{p}>0$ should 
be properly included).

\subsection{ Analysis of the global rate and highest energy bins}

We can make an estimation of the upper bound on the 
appearance solar electron antineutrino probability simply counting 
the number of observed events and subtracting the 
number of events expected from the best-fit oscillation 
solution. For our purposes this difference, which in this 
case is positive, can be interpreted as a hypothetical 
signal coming from  
solar antineutrinos
 ($\Delta m^2_0=6.9\times 10^5\ eV^2,
\sin^2 \theta_0=1)$.:
\begin{eqnarray}
S_{solar}=\overline{p}\times S_{solar}^0&=& 
S_{obs}-S_{react}(\Delta m^2_0,\sin^2 \theta_0).
\end{eqnarray}
Putting \cite{kloctober} 
$S_{obs}=54.3\pm 7.5 $ and
$S_{react}(\Delta m^2_0,\sin^2 \theta_0)= 49\pm 1.3 $,
 we obtain
$S_{obs}-S_{react}< 64.8\  (67.2)$ at 90 (95)\% CL.
From these numbers, the corresponding limits on 
solar electron antineutrino appearance probability are
$\overline{p}< 0.45\%, \ 0.52\% $ at 90 or 95\% CL. 
These limits are valid for the neutrino energy range
 $E_{\nu}\sim 2-8 $ MeV. 
In this case, due to the large range, the limits are 
better interpreted as limits on an 
energy-averaged probability
 according to expression \ref{avnu}. 

In a similar approach, we use on continuation 
the  binned KamLAND signal corresponding to 
the four highest energy bins
(see  Fig.\ref{f1}) which, as we will see, provide the strongest 
statistical significance and bounds. The reason for that is   
that the experiment KamLAND does not observe any signal here 
and, furthermore,
 the expected signal from oscillating neutrinos with LMA 
parameters is negligibly small.

Due to the small sample,
we apply Poisson statistics to any of these bins and use 
the fact that a sum of Poisson variables of mean $\mu_i$ is 
itself a Poisson variable of mean $\sum \mu_i$.
The background (here the reactor antineutrinos) and the 
signal (solar electron antineutrinos) are assumed  to be independent
Poisson Random variables with known means. 

If no events are 
observed, and, in particular, no background is observed, the 
unified intervals \cite{cousins,PDG2002}
$[0,\epsilon_{CL}]$ 
 are 
$[0,2.44]$ 
at 90\% CL and 
$[0,3.09]$ at 95\% CL.
From here, we obtain 
$\overline{p}\times S_0^{solar} < \epsilon_{CL}$ or 
$\overline{p}  < \epsilon_{CL}/S_0^{solar}$. Taking the 
expected number of events in the first 145 days of data taking 
and in this energy range (6-8 MeV) we obtain:
$\overline{p}<0.12 \% $ (90\% CL) and 
$\overline{p}<0.15\% $ (95\% CL).

\section{A model for solar antineutrino production}

\begin{figure}
\centering
\psfig{file=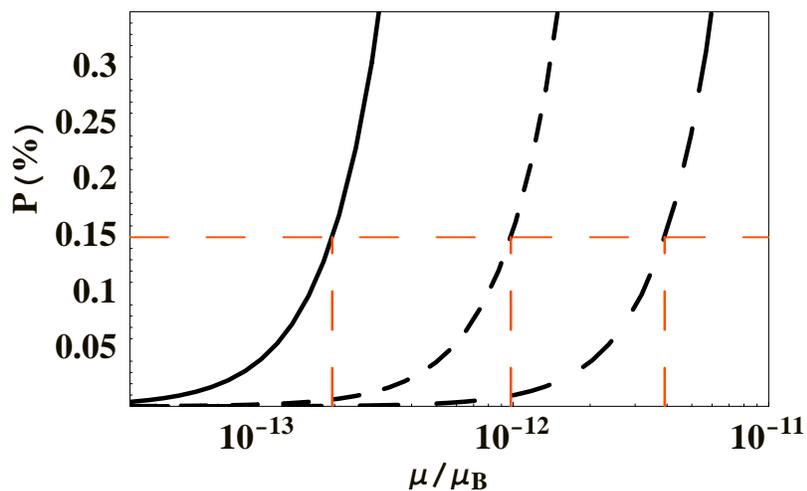,width=11cm} 
\caption{The solar antineutrino appearance probability $\overline{p}$ 
as a function of the transition neutrino magnetic moment, in units of 
Bohr magnetons $\mu_B$, for fixed values of the r.m.s solar magnetic 
field (Formula 
\protect\ref{anuprob}). From left (solid) to right (dashed), 
curves correspond to 
$B=1000,200,50$ kG. From the curves, an upper 
limit $\overline{p}<0.15\%$ implies 
$\mu< 1.9\times 10^{-13} \mu_B, 
9.0\times 10^{-13} \mu_B, 
3.0\times 10^{-12} \mu_B$ respectively for each of the magnetic 
field above. 
}
\label{f2}
\end{figure}

The combined action of  spin flavor precession in a 
magnetic field and  ordinary neutrino matter oscillations 
can produce an observable flux of $\anueR$'s from the Sun 
in the case of the neutrino being  a Majorana particle.
In the simplest model, where a thin layer of highly chaotic of 
magnetic field is assumed at the bottom of the convective 
zone (situated at $R\sim 0.7 R_\odot$), 
the antineutrino appearance 
 probability at the exit of the 
layer $P(\overline{\nu})$  is basically equal to 
the appearance probability of antineutrinos 
at the earth \cite{specificrandom,generalrandom} ( see also 
Refs.\cite{pulido} for some recent studies on RSFP solutions to
the Solar Neutrino Problem). 
The quantity  $P(\overline{\nu})$ is in general a function 
of the neutrino oscillation parameters  
$(\Delta m^2,\theta)$, the neutrino intrinsic magnetic moment
and also of the neutrino energy and 
the characteristics and magnitude of the solar 
magnetic field. However, in a accurate enough approximation, 
such probability can be factorized in a term depending 
only on the oscillation parameters and another one depending 
only on the spin-flavor precession parameters: 
\begin{equation}
P(\overline{\nu} )  = 
 \frac{1}{2}P_{e\mu}(\Delta m^2,\theta)\times 
\left [1-\exp\left (-4 \Omega^2 \Delta r  \right )  \right ]  
\label{anuprob}
\end{equation}
where $P_{e\mu}$ is the $e-\mu$ solar 
conversion probability. We will assume in this work the 
LMA central values for $(\Delta m^2,\theta)$ obtained 
from recent KamLAND data and which are compatible 
with the SNO observations in solar neutrinos \cite{alianidecember}, we will take 
$P_{e\mu}(\Delta m^2,\theta)\simeq
\langle P_{e\mu}\rangle^{exp,SNO} \simeq  0.4$. 
The second factor appearing in the expression 
contains  the 
effect of the magnetic field. 
This quantity depends on  the layer
width $\Delta r$  ($\sim 0.1 R_\odot$) and 
$\Omega^2\ \equiv  \frac{1}{3} L_0 \mu^2 \langle B^2\rangle$,
where
$\langle B^2\rangle$ 
the r.m.s strength of the
magnetic field 
and $L_0$ is a scale length ($L_0\sim 1000$ km).
For small values of the argument we have the following 
approximate expression 
which is accurate enough for many applications
$$P(\overline{\nu}) \simeq P_{e\mu}\times 2 \Omega^2 \Delta r=\kappa\ \mu^2 \langle B^2\rangle$$
the solar astrophysical  factor 
$\kappa\equiv 2/3 P_{e\mu} L_0 \Delta r$ is numerically 
$\kappa^{LMA}\simeq 2.8\times 10^{-44}$ MeV$^{-2}$.
Upper limits on the antineutrino appearance probability 
can be translated into upper limits 
 on the neutrino transition magnetic moment and the magnitude 
of the magnetic field in the solar interior.
The results of the Formula \ref{anuprob} 
can be seen in Figure \ref{f2}.
An upper  bound $\overline{p}< 0.15-0.20\%$ (95\% CL)
implies an upper limit on 
the product of the intrinsic neutrino magnetic moment and
the value of the convective solar magnetic field as 
$\mu B< 2.3\times  10^{-21}$ MeV (95\% CL).
In Fig.\ref{f2} we show the antineutrino probability 
as a function 
of the magnetic moment $\mu$ for fixed values of 
the magnitude of 
the magnetic field.
For realistic values of other astrophysical solar
 parameters ($L_0\sim 1000 $ km, $\Delta r\sim 0.1\ R_\odot$), 
these upper 
limits would imply that the neutrino magnetic 
moment is constrained to be, in the most desfavourable case, 
$\mu\lsim 3.9\times 10^{-12}\ \mu_B$ (95\% CL) for a relatively small
field $B= 50$ kG.  
Stronger limits are obtained for slightly 
higher values of the magnetic field:
 $\mu\lsim 9.0\times 10^{-13}\ \mu_B$ (95\% CL) for 
field $B= 200$ kG and 
 $\mu\lsim 2.0\times 10^{-13}\ \mu_B$ (95\% CL) for 
field $B= 1000$ kG. Let us note that these assumed values for 
the  magnetic field at the base the solar convective zone 
are relatively mild and well within present astrophysical 
expectatives.

\section{  Conclusions}
\label{sec:conclusions}

In summary in this work we investigate  the possibility of 
detecting 
solar antineutrinos with the KamLAND experiment.
These antineutrinos are predicted by spin-flavor solutions
to the solar neutrino problem.

The KamLAND experiment is  sensitive 
to potential 
antineutrinos originated from solar ${}^8$B neutrinos.
We find that the 
results of the KamLAND experiment put  
relatively strict limits on the flux of solar electron antineutrinos
$\Phi( {}^8 B)< 1.1-3.5\times 10^4\ cm^{-2}\ s^{-1}$,
and their energy averaged appearance probability ($P<0.15-0.50\%$).
These limits are largely independent from any model on the solar 
magnetic field or any other astrophysical properties.
As we remarked in Section 2.1, these  upper 
limits on antineutrino probabilities and fluxes 
are still 
valid even if the antineutrino probabilities are significantly 
different from constant.

Next we assume a concrete model for antineutrino production
where they are produced  
by spin-flavor precession in the convective solar 
magnetic field. In this model, the antineutrino  
appearance probability is given by a simple expression as
$P(\overline{\nu})=\kappa\ \mu^2 \langle B^2\rangle$
with  $\kappa^{LMA}\simeq 2.8\times 10^{-44}$ MeV$^{-2}$.
In the context of this model and
assuming LMA central values for neutrino oscillation 
parameters
($\Delta m^2=6.9\times 10^{-5}$ eV$^2$, $\sin^2\theta=1$) 
\cite{kloctober}, the  upper 
 bound $\overline{p}< 0.15\%$ (95\% CL)
implies an upper limit on 
the product of the intrinsic neutrino magnetic moment and
the value of the convective solar magnetic field as 
$\mu\ B< 2.3\times  10^{-21}$ MeV (95\% CL).
For realistic values of other astrophysical solar
 parameters these upper 
limits would imply that the neutrino magnetic 
moment is constrained to be, in the most desfavourable case, 
$\mu\lsim 3.9\times 10^{-12}\ \mu_B$ (95\% CL) for a relatively small field $B= 50$ kG.  
For slightly higher values of the magnetic field:
 $\mu\lsim 9.0\times 10^{-13}\ \mu_B$ (95\% CL) for 
field $B= 200$ kG and 
 $\mu\lsim 2.0\times 10^{-13}\ \mu_B$ (95\% CL) for 
field $B= 1000$ kG. 
These assumed values for 
the  magnetic field at the base the solar convective zone 
are relatively mild and well within present astrophysical 
expectatives.

\vspace{0.3cm}
\subsection*{Acknowledgments}

I would like to  acknowledge many useful conversations with 
P. Aliani, M. Picariello and V. Antonelli.
I  acknowledge the  financial  support of 
  the  Spanish CYCIT  funding agency.

\newpage

\end{document}